\documentstyle[12pt,epsfig]{article}
\thicklines

\begin{document}

\title{
%\begin{flushright} \large
%ATLAS Internal Note \\
%PHYS-NO-115 \\
%5 November 1997
%\end{flushright}
%\vspace{1.5cm}
\Large
$B_s^0 \to D_s^-a_1^+(D_s^- \to \phi \pi^-,~D_s^- \to K^{*0}K^-)$ decay 
channel in the ATLAS $B_s^0$-mixing studies}
\protect\author{\fbox{A.V.~Bannikov}, G.A.~Chelkov, Z.K.~Silagadze\thanks
{ permanent address: Budker Institute of Nuclear Physics,
630 090, Novosibirsk, Russia.}
\vspace*{3mm} \\
\sc JINR, Dubna \vspace*{3mm} \\}
\date{}
\maketitle

\begin{abstract}
It is shown, using a track-level simulation, that the use of the
$D_s^- \to K^{*0}K^-$ decay channel for $D_s^-$ reconstruction,
in addition with the previously studied $D_s^- \to \phi \pi^-$
mode, enables two fold gain in the ATLAS $B_s^0$-mixing signal
statistics through $B_s^0 \to D_s^-a_1$ ~$B_s^0$-decay channel.
A new modification of the amplitude fit method is suggested for
the $x_s$ determination.
Some general aspects of the $B_s^0$-mixing phenomenon is illustrated
by pictures of Casimir Malevich, Maurits Cornelis Escher and Salvador 
Dali.
\end{abstract}

\section{Introduction}
The main features of the $B_s$-mixing studies were explained
in previous notes \cite{1,2} (see also references cited therein),
so we don't need to repeat them here. Instead we will try to give
arguments that $B_s$-mixing phenomenon is indeed worthy to be studied.

The famous Russian painter Casimir Malevich said a long time ago:
"The object in itself is meaningless ... the ideas of the conscious
mind are worthless". We would like to choose his great painting
"The black square", which is reproduced below, as a starting point
of our introduction.

\unitlength=1mm
\begin{picture}(100,30)
\put(45,28){\line(2,0){25}}
\put(70,28){\line(0,-2){25}}
\put(70,3){\line(-2,0){25}}
\put(45,3){\line(0,2){25}}
\linethickness{15mm}
\put(50,15.5){\line(2,0){15}}
\end{picture}

But from this starting point it is possible to go to the very different
directions depending from one's imagination. So let us imagine the
following picture behind the black square \cite{3} :

"A cat is penned up in a steel chamber, along with the following
diabolical device: in a Geiger counter there is a tiny bit of radioactive
substance, that perhaps in the course of one hour one of the atoms decays,
but also, with equal probability, perhaps none; if it happens, the counter
tube discharges and through a relay releases a hammer which shatters
a small flask of hydrocyanic acid. If one has left the entire system
to itself for one hour, one would say that the cat still lives if
meanwhile no atom has decayed. The state vector $|\Psi>$ of the entire
system would express this by having in it the living and the dead cat
mixed or smeared out in equal parts."

But this is of course nonsense, at least from cat's point of view!

We have reminded Schr\"{o}dinger's cat old story here in order to give
an impression that although we all became familiar with particle mixing,
because the superposition principle lies on a very background of quantum
mechanics, this phenomenon is by no means obvious or trivial property of
reality.

But what is strange and queer at the macrophysics level can still appear
as the most common thing at the microphysics level. It seems that even our
existence is based on particle mixing as will be explained below.

One of very important characteristics of elementary particle is its mass.
We can get some insight about its origin from the following simple trick.
The propagator of a massive fermion can be represented in such a way
\begin{eqnarray}
\frac{1}{\hat{p}-m}=\frac{1}{\hat{p}}+\frac{1}{\hat{p}}m\frac{1}{\hat{p}}+
\frac{1}{\hat{p}}m\frac{1}{\hat{p}}m\frac{1}{\hat{p}}+\cdots
\; ,
\nonumber \end{eqnarray}
or graphically

\begin{picture}(120,10)
\put(15,5){\line(2,0){5}}
\put(21.5,5){\circle*{3}}
\put(23,5){\line(2,0){5}}
\put(30,5.5){\line(2,0){2}}
\put(30,4.5){\line(2,0){2}}
\put(34,5){\line(2,0){5}}
\put(41,4){+}
\put(46,5){\line(2,0){17}}
\put(52.5,6){\line(2,-1){4}}
\put(52.5,4){\line(2,1){4}}
\put(53,8){m}
\put(65,4){+}
\put(70,5){\line(2,0){17}}
\put(76.5,6){\line(2,-1){4}}
\put(76.5,4){\line(2,1){4}}
\put(77,8){m}
\put(87,5){\line(2,0){17}}
\put(93.5,6){\line(2,-1){4}}
\put(93.5,4){\line(2,1){4}}
\put(94,8){m}
\put(106,4){$\cdots$}
\end{picture}

where a single line represents the propagator of a massless particle.
So things look like as if the massless particle is propagating through
some medium and the mass emerges as a result of friction or interaction
with this environment. But what is the medium the particle interacts with?
A (massless) fermionic particle can have the following interaction with
some scalar field ${\cal{L}}_{int}=g \bar \psi \psi \varphi$:

\begin{picture}(120,20)
\put(50,2){\vector(1,2){2}}
\put(52,6){\line(1,2){2}}
\put(54,10){\vector(-1,2){2}}
\put(52,14){\line(-1,2){2}}
\put(54,10){\circle*{1}}
\put(54,10){\line(2,0){2}}
\put(58,10){\line(2,0){2}}
\put(62,10){\line(2,0){2}}
\put(66,10){\line(2,0){2}}
\put(70,9){$\varphi$}
\put(50,9){$g$}
\end{picture}

If now the self interactions of this scalar field are such that it doesn't
disappear in a vacuum state and develops a nonzero vacuum expectation value
$<\varphi>$, when it is convenient to expand $\varphi=<\varphi>+
\varphi^\prime$, where $\varphi^\prime$ corresponds to the physical scalar
particles (excitations over the vacuum) and $<\varphi>$ just gives the
medium (the vacuum) where all of us
are living. Now because of this decomposition of $\varphi$ the fermion-scalar
interaction splits into two parts:

\begin{picture}(125,20)
\put(10,2){\vector(1,2){2}}
\put(12,6){\line(1,2){2}}
\put(14,10){\vector(-1,2){2}}
\put(12,14){\line(-1,2){2}}
\put(14,10){\circle*{1}}
\put(14,10){\line(2,0){2}}
\put(18,10){\line(2,0){2}}
\put(22,10){\line(2,0){2}}
\put(26,10){\line(2,0){2}}
\put(30,9){$\varphi$}
\put(10,9){$g$}
\put(40,10){\vector(2,0){10}}
\put(55,2){\vector(1,2){2}}
\put(57,6){\line(1,2){2}}
\put(59,10){\vector(-1,2){2}}
\put(57,14){\line(-1,2){2}}
\put(59,10){\circle*{1}}
\put(57,11){\line(2,-1){4}}
\put(57,9){\line(2,1){4}}
\put(62,9){$g<\varphi>$}
\put(83,9){$+$}
\put(95,2){\vector(1,2){2}}
\put(97,6){\line(1,2){2}}
\put(99,10){\vector(-1,2){2}}
\put(97,14){\line(-1,2){2}}
\put(99,10){\circle*{1}}
\put(99,10){\line(2,0){2}}
\put(103,10){\line(2,0){2}}
\put(107,10){\line(2,0){2}}
\put(111,10){\line(2,0){2}}
\put(115,9){$\varphi^\prime$}
\put(95,9){$g$}
\end{picture}

The second diagram represents an emission of the real scalar quantum and
the first one generates the fermion mass $m=g<\varphi>$.

But if, for example $d$-quark can emit a scalar particle without changing
its flavour, why can't it do this with changing the flavour? We know that
the flavour is not always conserved, so the following interaction is not
excluded:

\begin{picture}(125,20)
\put(10,2){\vector(1,2){2}}
\put(12,6){\line(1,2){2}}
\put(14,10){\vector(-1,2){2}}
\put(12,14){\line(-1,2){2}}
\put(14,10){\circle*{1}}
\put(6,2){$d$}
\put(6,16){$s$}
\put(14,10){\line(2,0){2}}
\put(18,10){\line(2,0){2}}
\put(22,10){\line(2,0){2}}
\put(26,10){\line(2,0){2}}
\put(30,9){$\varphi$}
\put(10,9){$g$}
\put(38,10){\vector(2,0){10}}
\put(55,2){\vector(1,2){2}}
\put(57,6){\line(1,2){2}}
\put(59,10){\vector(-1,2){2}}
\put(57,14){\line(-1,2){2}}
\put(59,10){\circle*{1}}
\put(51,2){$d$}
\put(51,16){$s$}
\put(57,11){\line(2,-1){4}}
\put(57,9){\line(2,1){4}}
\put(62,9){$g<\varphi>$}
\put(83,9){$+$}
\put(95,2){\vector(1,2){2}}
\put(97,6){\line(1,2){2}}
\put(99,10){\vector(-1,2){2}}
\put(97,14){\line(-1,2){2}}
\put(99,10){\circle*{1}}
\put(91,2){$d$}
\put(91,16){$s$}
\put(99,10){\line(2,0){2}}
\put(103,10){\line(2,0){2}}
\put(107,10){\line(2,0){2}}
\put(111,10){\line(2,0){2}}
\put(115,9){$\varphi^\prime$}
\put(95,9){$g$}
\end{picture}

But now the first term gives $d-s$ mixing! As a result our initial $d$
and $s$ fields are no longer mass eigenstates (the states with definite
mass), instead their time development in the rest frame is described by
the matrix Schr\"{o}dinger's equation ($\hbar=1$):
$$
i\frac{\partial}{\partial t}
\left ( \begin{array}{c} d \\ s  \end{array} \right ) =
\left ( \begin{array}{cc} m_d & m_{ds} \\ m_{ds} & m_s \end{array} \right )
\left ( \begin{array}{c} d \\ s  \end{array} \right )
\equiv $$ $$ \left ( \begin{array}{cc} \cos{\theta_c} & \sin{\theta_c} \\
-\sin{\theta_c} & \cos{\theta_c} \end{array} \right )
\left ( \begin{array}{cc} \tilde{m}_d & 0 \\ 0 & \tilde{m}_s \end{array}
\right ) \left ( \begin{array}{cc} \cos{\theta_c} & -\sin{\theta_c} \\
\sin{\theta_c} & \cos{\theta_c} \end{array} \right )
\left ( \begin{array}{c} d \\ s  \end{array} \right )
\; , $$ \noindent
where $\tan{2\theta_c}=\frac{2m_{ds}}{m_s-m_d}$ and $\tilde{m}_d \; , \;
\tilde{m}_s$ mass eigenvalues are defined from the equations $m_d=
\tilde{m}_d \cos^2{\theta_c}+\tilde{m}_s \sin^2{\theta_c} \; , \;
m_s=\tilde{m}_d \sin^2{\theta_c}+\tilde{m}_s \cos^2{\theta_c}$. It is
obvious that the corresponding eigenvectors (the physical $d$ and $s$
quarks) are
$$\left ( \begin{array}{c} \tilde{d} \\ \tilde{s}  \end{array} \right ) =
\left ( \begin{array}{cc} \cos{\theta_c} & -\sin{\theta_c} \\
\sin{\theta_c} & \cos{\theta_c} \end{array} \right )
\left ( \begin{array}{c} d \\ s  \end{array} \right ) \longrightarrow
\begin{array}{c} \tilde{d}=\cos{\theta_c}~d-\sin{\theta_c}~s \\
\tilde{s}=\sin{\theta_c}~d+\cos{\theta_c}~s \end{array} $$

This particle mixing has one important observable consequence. If initially
the weak transitions were possible only within the $(u,d)$ or $(c,s)$ pairs,
now the intergeneration transitions $\tilde{u} \leftrightarrow \tilde{s}$
and $\tilde{c} \leftrightarrow \tilde{d}$ are also possible because, for
example, the physical $s$-quark contains both $d$ and $s$ bare fields
$\tilde{s}=\sin{\theta_c}d+\cos{\theta_c}s$. Thus $\tilde{u} \rightarrow
\tilde{s}$ transition is proportional to $\sin{\theta_c}$ - sine of the so
called Cabibo angle.

But we have three quark-lepton generations. So after the mixing the weak
transitions are possible between any up and any down quarks. The amplitudes
of these weak transitions are convenient to express as a $3 \times 3$
unitary matrix. This Kobayashi-Maskawa matrix is a generalization of the
Cabibo angle and reveals a remarkable hierarchical structure \cite{4}
\begin{eqnarray}
\left ( \begin{array}{ccc} V_{ud} & V_{us} & V_{ub} \\ V_{cd} & V_{cs}
& V_{cb} \\ V_{td} & V_{ts} & V_{tb} \end{array} \right ) \approx
\left ( \begin{array}{ccc} 1-\frac{1}{2}\lambda^2 & \lambda &
A\lambda^3(\rho-i\eta) \\ -\lambda & 1-\frac{1}{2}\lambda^2
& A\lambda^2 \\ A\lambda^3(1-\rho-i\eta) & -A\lambda^2 & 1 \end{array}
\right ) \; .
\label{eq1} \end{eqnarray}
Here $\lambda=\sin{\theta_c}\approx0.22$ is a small quantity. So the
intergeneration weak transitions are suppressed and this suppression is
more strong for not neighboring generations.

If $\eta \neq 0$, the Kobayashi-Maskawa matrix is complex and violates $CP$.
It is commonly believed today that this $CP$-violation is an important
ingredient in baryon-antibaryon asymmetry generation in the universe \cite{5}
and so the source of our very existence.

So far we were talking about particle mixing at quark level. But quarks are
confined into hadrons and we can study quark-mixing only indirectly via
hadron-mixing. $B$-meson system is very promising in this respect: because
of a large mass of the $b$-quark we can enjoy an asymptotic freedom advantage
of QCD and calculate strong interaction corrections, unlike, for example,
$K$-meson system.

In the Standard Model the $B_d-\bar{B}_d$ mixing originates from the
following diagram (and from the
second one there intermediate up-quark and $W$ lines are interchanged)

\begin{picture}(125,40)
\put(17,29){$b$}
\put(23,10){\line(2,0){10}}
\put(43,10){\vector(-2,0){10}}
\put(43,10){\line(0,2){10}}
\put(43,30){\vector(0,-2){10}}
\put(23,30){\vector(2,0){10}}
\put(33,30){\line(2,0){10}}
\put(17,9){$\bar d$}
\put(43,10){\circle*{1}}
\put(43,30){\circle*{1}}
\put(43,10){\line(1,-2){2}}
\put(45,6){\line(1,2){3}}
\put(48,12){\line(1,-2){3}}
\put(51,6){\line(1,2){3}}
\put(54,12){\line(1,-2){3}}
\put(57,6){\line(1,2){3}}
\put(60,12){\line(1,-2){3}}
\put(63,6){\line(1,2){2}}
\put(37,19){$u_i$}
\put(43,30){\line(1,2){2}}
\put(45,34){\line(1,-2){3}}
\put(48,28){\line(1,2){3}}
\put(51,34){\line(1,-2){3}}
\put(54,28){\line(1,2){3}}
\put(57,34){\line(1,-2){3}}
\put(60,28){\line(1,2){3}}
\put(63,34){\line(1,-2){2}}
\put(68,19){$u_j$}
\put(85,29){$\bar b$}
\put(85,9){$d$}
\put(65,10){\vector(2,0){10}}
\put(75,10){\line(2,0){10}}
\put(65,10){\line(0,2){10}}
\put(65,30){\vector(0,-2){10}}
\put(85,30){\vector(-2,0){10}}
\put(65,30){\line(2,0){10}}
\put(65,10){\circle*{1}}
\put(65,30){\circle*{1}}
\put(52,24){$W$}
\put(52,13){$W$}
\put(39,33){$V_{tb}$}
\put(39,5){$V^*_{td}$}
\put(65,5){$V^*_{td}$}
\put(65,33){$V_{tb}$}
\end{picture}

\noindent $u_i$ stands for any up quark. So
$$B_d-{\rm mixing} \sim \sum_{i,j} \lambda_i \lambda_j I(m_i,m_j)$$
\noindent where $\lambda_i=V_{u_ib}V^*_{u_id}$ and $I(m_i,m_j)$ represents
the loop integral. This integral diverges quadratically. But this divergence
is harmless because the unitarity of the Kobayashi-Maskawa matrix ensures
its cancellation in the sum: the unitarity means $\sum \lambda_i=0$,
therefore
$$\sum_{i,j} \lambda_i \lambda_j I(m_i,m_j)=
\sum_{i,j} \lambda_i \lambda_j [ I(m_i,m_j)-I(0,m_j)-I(m_i,0)+I(0,0) ] $$
\noindent and these subtractions greatly improve the convergence. For
example, in case of $t$-quark contribution, these subtractions lead to the
replacement
$$\frac{1}{(k^2-m_t^2)^2} \longrightarrow \frac{1}{(k^2-m_t^2)^2} -
\frac{2}{k^2(k^2-m_t^2)} + \frac{1}{k^2 k^2}=\frac{m_t^4}{k^2 k^2(k^2-m_t^2)}
\; . $$

From this expression it is also clear that in fact just $t$-quark
contribution is dominant for $B$-mixing, because of its extraordinary large
mass.

When ARGUS made his measurement of the $B_d$-mixing \cite{6}, nobody thought
that $t$-quark is so massive. So the result of this measurement appeared as
a big surprise. We can even say that $t$-quark was discovered by ARGUS,
because the large $B_d$-mixing, observed by ARGUS, is very difficult to
explain without the existence of the $t$-quark with mass $>100 GeV$.

We can infer from the above given diagram that even larger mixing is
expected in $B_s$-system:
$$\frac{B_s-{\rm mixing}}{B_d-{\rm mixing}} \sim \left | \frac{V_{ts}}
{V_{td}} \right |^2 \sim \frac{1}{\lambda^2[(1-\rho)^2+\eta^2]} \sim
\frac{1}{\lambda^2} \sim 25 $$
We see also that the relative magnitude of $B_s$ and $B_d$ mixings measures
$(1-\rho)^2+\eta^2$ - one side of the notorious unitarity triangle \cite{7}.
It is worthwhile to mention that this ratio is, to a great extent, free from
hadronic uncertainties, which arise when we ask how quark and antiquark from
the above given $B$-mixing diagram really form $B$-meson.

To summarize, the $B_s$-mixing studies are interesting, because they reveal
a very fundamental underlying phenomenon - the generation of particle masses
and mixing angles via the Higs mechanism, the least understood thing in the
Standard Model. Because of heaviness of the $b$-quark and asymptotic freedom
of QCD, the theory gives very definite predictions about expected
$B_s$-mixing, hampered only from uncertainties due to our inability to solve
QCD in the confinement region. But these uncertainties are also, to a certain
extent, under control \cite{7}. The theoretical predictions involve such a
fundamental property as the unitarity of the Kobayashi-Maskawa matrix
(the existence of only three generations). Any deviation between the theory
and experiment can lead to significant change of our present day picture
of the elementary particle world (recall the $B_d$-mixing story). The
forthcoming ATLAS experiment sensitivity to the $B_s$-mixing covers the
Standard Model prediction range \cite{8}. So it will either give one more
conformation of the theory or will open a window into a physics beyond
the Standard Model.

\section{Black Square view on the $B_s$-mixing}
We hope the above given considerations convinced the reader that the
Black Square can hide a very reach content behind it. For example,
reflecting about the Schr\"{o}dinger's cat we can end with the following
picture of $B_d$-oscillations (experimentally confirmed by ARGUS 
\cite{6})

% !!!!!!!!!!!!!!!!!!!!!!!!!!!!!!!!!!!!!!!!!!!!!!!!!!!!!!!
% escher1.ps is too big. I'm sending escher1.jpg instead.
\vspace*{5cm}
\centerline{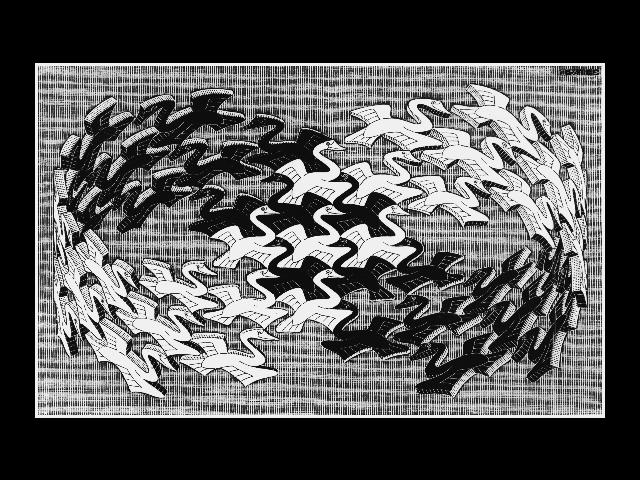 here}
\vspace*{7cm}
% !!!!!!!!!!!!!!!!!!!!!!!!!!!!!!!!!!!!!!!!!!!!!!!!!!!!!!!
% 
%\begin{figure}[htb] 
%  \begin{center}    
%    \mbox{\epsfig{figure=escher1.ps,% 
%                        height=9.0cm}}
%  \end{center}      
%\end{figure}
\newpage
As we have already mentioned, more rapid oscillations are expected in the
$B_s$ system. So if man eagerly stares on the Black Square he will at last 
distinguish the picture of $B_s$-mixing:

% !!!!!!!!!!!!!!!!!!!!!!!!!!!!!!!!!!!!!!!!!!!!!!!!!!!!!!!
% escher2.ps is too big. I'm sending escher2.jpg instead.
\vspace*{4cm}
\centerline{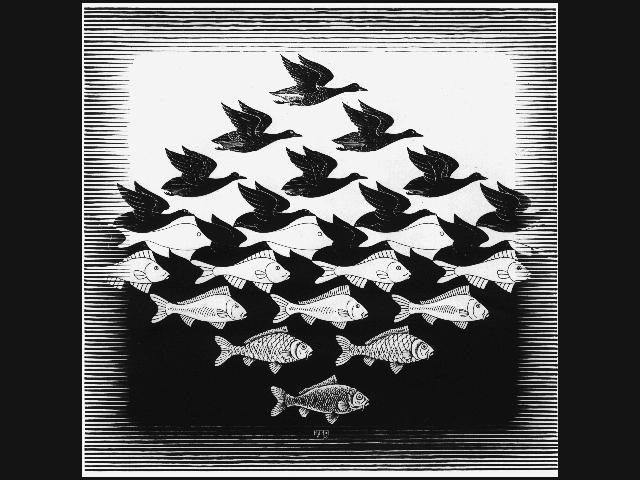 here}
\vspace*{4cm}
% !!!!!!!!!!!!!!!!!!!!!!!!!!!!!!!!!!!!!!!!!!!!!!!!!!!!!!!
%
%\begin{figure}[htb]
%  \begin{center}
%    \mbox{\epsfig{figure=escher2.ps,%
%                        height=9.0cm}}
%  \end{center}
%\end{figure}

The most impressive people can remark even a symbolization of the 
$CP$-violation on this illustration: the nature for some reason makes 
an absolute difference between particle and antiparticle, not just the 
conventional difference between black and white.

\section{$D_s^- \to K^{*0}K^-$ decay channel for $D_s$ reconstruction}
To observe the $B_s^0$-mixing in a real experiment like ATLAS and extract
the corresponding $x_s$ parameter, which characterizes the $B_s-\bar{B_s}$
oscillation frequency, you need to reconstruct $B_s$ meson and determine
its decay vertex with great precision. Two of the $B_s$ decay channels were
considered for this goal up to now: $B_s^0 \to D_s^- \pi^+$ \cite{1} and
$B_s^0 \to D_s^- a_1^+$ \cite{2}. For the second channel $D_s^- \to \phi
\pi^-, ~\phi \to K^+K^-$ decay mode was used for the $D_s^-$ reconstruction.
It was mentioned in \cite{1,2} that other decay channels of $D_s^-$ can be
also used to increase signal statistics. In the present note we consider
$D_s^- \to K^{*0} K^-, ~  K^{*0} \to K^+ \pi^-$ decay mode as one of the 
possibilities:

\begin{picture}(150,110)
\put(20,40){\circle*{4}}
\put(20,40){\vector(-3,-2){30}}
\put(-8,17){{\Large\bf $\mu_{tag}$}}
\put(20,40){\line(2,1){40}}
\put(42,45){{\Large\bf $B^0_s$}}
\put(60,60){\line(3,-1){15}}
\put(63,52){{\Large\bf $D^-_s$}}
\put(75,55){\vector(3,1){33}}
\put(110,65){{\Large\bf $K^-$}}
\put(75,55){\vector(3,0){35}}
\put(113,55){{\Large\bf $K^+$}}
\put(75,55){\vector(3,-2){35}}
\put(110,35){{\Large\bf $\pi^-$}}
\put(60,60){\vector(1,3){10}}
\put(65,90){{\Large\bf $\pi^+$}}
\put(60,60){\vector(2,3){20}}
\put(82,90){{\Large\bf $\pi^-$}}
\put(60,60){\vector(3,2){45}}
\put(100,90){{\Large\bf $\pi^+$}}
\put(125,62){\oval(3,15)[r]}
\put(122,45){\oval(3,30)[r]}
\put(130,62){{\Large\bf $\phi$}}
\put(130,56){{\large $(1^{st}~mode)$}}
\put(127,45){{\Large\bf $K^{*0}$}}
\put(127,39){{\large $(2^{nd}~mode)$}}
\put(77,95){\oval(25,4)[t]}
\put(77,100){{\Large\bf $\rho^0$}}
\put(87,105){\oval(44,4)[t]}
\put(87,110){{\Large\bf $a^+_1$}}
\end{picture}

 As it is clear from the Table 1 below, this decay channel
is quite promising if compared to the previously used $D_s^- \to \phi 
\pi^-$.

%********** Table 1 ************
%\newpage
\begin{center}
Table 1.\\
Branching ratios and signal statistics
for $B_s^0\rightarrow D_s^{-}a_1^{+}(1260)$.\\
\vspace{0.5cm}
\begin{tabular}{|l|c|l|}
\hline
\hline
Parameter & Value & Comment \\
\hline
\hline
  $L~[cm^{-2}s^{-1}]$                          & $10^{33}$  &   \\
  $t~[s]$                                      & $10^7$     &   \\
%  $\sigma(b \bar b)/\sigma(tot)$               & $\simeq 1/100$ &   \\
%  $\sigma(b \bar b)~[\mu b]$                   & $\simeq 500$   &   \\
  $\sigma(b \bar b \rightarrow \mu X)~[\mu b]$ & $ 2.3$   &
  $p^{\mu}_{T} > 6~GeV/c$  \\
					       &            &
  $|\eta^{\mu}| < 2.2$     \\
\hline
  $N(b \bar b \rightarrow \mu X)$              & $2.3 \times 10^{10}$ &  \\
\hline
  $Br(b \rightarrow B^0_s)$                    & $0.112$    &   \\
  $Br(B^0_s \rightarrow D^-_{s}a_1^+)$         & $0.006$    &   \\
  $Br(a_1^+ \rightarrow \rho^0\pi^+)$          & $\sim 0.5$ &   \\
  $Br(\rho^0 \rightarrow \pi^-\pi^+)$          & $\sim 1$   &   \\
\hline
  $Br(D^-_s \rightarrow \phi\pi^-)$            & $0.036$    &   \\
  $Br(\phi \rightarrow K^+K^-)$                & $0.491$    &   \\
\hline
  $Br(D^-_s \rightarrow K^{*0}K^-)$            & $0.034$    &   \\
  $Br(K^{*0} \rightarrow K^+\pi^-)$            & $\sim 0.65$     &   \\
\hline
\hline
  $N(K^+K^-\pi^-\pi^+\pi^-\pi^+)$              & $136600$    &
$D^-_s \rightarrow \phi\pi^-$  \\
\hline
\hline
  $N(K^+K^-\pi^-\pi^+\pi^-\pi^+)$              & $170800$    &
$D^-_s \rightarrow K^{*0}K^-$  \\
\hline
\hline
\end{tabular}
\end{center}
\vspace{0.5cm}

%*******************************

Event generation and reconstruction procedures are similar to ones 
considered in \cite{2}. All other general parameters, like impact
parameter resolution for smearing and transverse momentum resolution,
are also the same as in \cite {2} and can be found there.

Contrary to the  $D_s^- \to \phi \pi^-$ case, where $D_s^-$ peak was 
clearly seen in the invariant mass distribution of the three properly 
charged particles, assuming that two of them are $K$-mesons and one is pion,
now $D_s^-$ peak is not seen (Fig. 1), perhaps because $K^{*0}$ is too wide as
compared to $\phi$. Thus the combinatorial background from the signal events
alone is already able to hide the $D_s$ and 
the reader is left with the sentence \cite{9}
"if you should see the word "buffalo" written on a cage containing an
elephant, don't believe your eyes". Although when $D_s^-$ meson is 
reconstructed from its true decay products the resulting invariant mass 
resolution, shown on 
Fig. 2, is almost the same as for the $D_s^- \to \phi \pi^-$ mode. The
another picture on Fig. 2 shows $K^{*0}$, reconstructed from its true decay
products.

The resolution in the $B_s$-decay proper time (Fig. 3a) $\sigma_\tau
\approx 0.061ps$ is practically the same as for the $D_s^- \to \phi \pi^-$
mode. The corresponding $B_s$-decay length resolution in the transverse
plane is $\approx 100\mu m$ and the relevant distribution is shown on 
Fig. 3b.

We expect that signal to background situation when using $D_s^- \to K^{*0}
K^-$ mode will be similar to what was considered in \cite{2}. Compare for
example Fig. 4 from \cite{2} and from this work, which describes a possible
background from the $B_d^0 \to D_s^- a_1^+$ decay when $D_s^-$ is
reconstructed via $D_s^- \to \phi \pi^-$ or $D_s^- \to K^{*0}K^-$ modes
respectively.

The main reason which allowed a good signal to background separation in
\cite{2} was the fact that $D^-$ and $B_d^0$ masses are shifted from the
$D_s^-$ and $B_s^0$ masses by about 100 $MeV$. But this equally applies to
the $D_s^- \to K^{*0}K^-$ case also, because our cut on the $D_s^-$
invariant mass doesn't change very much. The only problem which can arise
is a large $K^{*0}$ width (as compared to the $\phi$-meson width) and
therefore the cut on the invariant mass of $K^{*0}$ should be considerably
loose. At present we are not aware of a background for which this 
circumstance will play a crucial role.

Acceptance and analysis cuts are summarized in Table 2. As it is known 
\cite{10}, a second level trigger is necessary to reduce an event rate,
which is still too high after the first level trigger (the tag-muon).
For the $B_s \to D_s^- \pi^+, \; D_s^- \to \phi \pi^-$ mode the problem
was studied in \cite{10} and it was shown that some loose cuts on the
invariant masses of $\phi$ and $D_s$ candidates can be used for this
purpose. The resulting trigger efficiency appeared to be \cite{8} about
0.54. For the $B_s \to D_s^- a_1^+, \; D_s^- \to K^{*0}K^-$ channel,
discussed in this note, the second level trigger should be specially
investigated, of course. However we expect that the similar mass cuts
on $D_s$ and $K^{*0}$ candidates will work in this case also and
a $50\%$ trigger efficiency should be a safe estimate. 

\newpage
%********** Table 2 ************

\begin{center}
Table 2.\\
Number of signal events from $B_s^0\rightarrow D_s^{-}a_1^{+}(1260)$
channel expected in ATLAS after 1 year ($10^7 s$) of operation at
$10^{33} cm^{-2}s^{-1}$.\\
\vspace{0.5cm}
\begin{tabular}{|l|c|l|}
\hline
\hline
Parameter & Value & Comment \\
\hline
\hline
  $N(K^+K^-\pi^-\pi^+\pi^-\pi^+)$              & $136600$    &
$D^-_s \rightarrow \phi\pi^-$ \\
\hline
  $N(K^+K^-\pi^-\pi^+\pi^-\pi^+)$              & $170800$    &
$D^-_s \rightarrow K^{*0}K^-$ \\
\hline
  Cuts :                         &            &   \\
  $p_{T} > 1~GeV/c$                            &            &   \\
  $|\eta| < 2.5$                               &            &   \\
\hline
  $N(K^+K^-\pi^-\pi^+\pi^-\pi^+)$              & $9015~~(6.6\%)$     &
$D^-_s \rightarrow \phi\pi^-$ \\
\hline
  $N(K^+K^-\pi^-\pi^+\pi^-\pi^+)$              & $9910~~(5.8\%)$     &
$D^-_s \rightarrow K^{*0}K^-$ \\
\hline
  $\Delta\varphi_{\pi\pi} < 35^\circ$          &            &   \\
  $\Delta\theta_{\pi\pi} < 15^\circ$           &            &   \\
  $|M_{\pi\pi}-M_{\rho^0}| < 192~MeV/c^2~(\pm 3\sigma) $    &     &   \\
  $|M_{\pi\pi\pi}-M_{a_1^+}| < 300~MeV/c^2$    &            &   \\
\hline
  $\Delta\varphi_{KK} < 10^\circ$              &            &
$D^-_s \rightarrow \phi\pi^-$ \\
  $\Delta\theta_{KK} < 10^\circ$               &            &   \\
  $|M_{KK}-M_{\phi}| < 20~MeV/c^2$             &            &   \\
  $|M_{KK\pi}-M_{D_s^-}| < 15~MeV/c^2$         &            &   \\
\hline
  $\Delta\varphi_{K\pi} < 20^\circ$            &            &
$D^-_s \rightarrow K^{*0}K^-$ \\
  $\Delta\theta_{K\pi} < 10^\circ$             &            &   \\
  $|M_{K\pi}-M_{K^{*0}}| < 80~MeV/c^2$         &            &   \\
  $|M_{KK\pi}-M_{D_s^-}| < 20~MeV/c^2$         &            &   \\
\hline
  $N(K^+K^-\pi^-\pi^+\pi^-\pi^+)$              & $6830~~(5.0\%)$     &
$D^-_s \rightarrow \phi\pi^-$ \\
\hline
  $N(K^+K^-\pi^-\pi^+\pi^-\pi^+)$              & $6830~~(4.0\%)$     &
$D^-_s \rightarrow K^{*0}K^-$ \\
\hline
  $D_s^{-}$ vertex fit $\chi^2 < 12.0$         &            &   \\
  $a_1^{+}$ vertex fit $\chi^2 < 12.0$         &            &   \\
%  $B_s^{0}$ vertex fit $\chi^2 < 0.35$         &            &   \\
  $B_s^{0}$ proper decay time $> 0.4~ps$       &            &   \\
  $B_s^{0}$ impact parameter $< 55~\mu m$      &            &   \\
  $B^0_s~~p_T > 10.0~GeV/c$                    &            &   \\
\hline
  $N(K^+K^-\pi^-\pi^+\pi^-\pi^+)$ after cuts   & $4100~~(3.0\%)$     &
$D^-_s \rightarrow \phi\pi^-$ \\
\hline
  $N(K^+K^-\pi^-\pi^+\pi^-\pi^+)$ after cuts   & $4780~~(2.8\%)$     &
$D^-_s \rightarrow K^{*0}K^-$ \\
\hline
  Lepton identification                        & $0.8$      &   \\
  Hadron identification                        & $(0.95)^6$ &   \\
  Trigger efficiency                           & $0.5$     &   \\
%\hline
%  $N(K^+K^-\pi^-\pi^+\pi^-\pi^+)$              & $1310~~(1.0\%)$   &
%$D^-_s \rightarrow \phi\pi^-$ \\
%\hline
%  $N(K^+K^-\pi^-\pi^+\pi^-\pi^+)$              & $1460~~(0.9\%)$   &
%$D^-_s \rightarrow K^{*0}K^-$ \\
%\hline
  Mass cut $\pm 2\sigma$                       &  $0.95$            &     \\
\hline
\hline
  $N(K^+K^-\pi^-\pi^+\pi^-\pi^+)$ reconstructed  & $1240~~(0.9\%)$   &
$D^-_s \rightarrow \phi\pi^-$ \\
\hline
\hline
  $N(K^+K^-\pi^-\pi^+\pi^-\pi^+)$ reconstructed  & $1330~~(0.8\%)$   &
$D^-_s \rightarrow K^{*0}K^-$ \\
\hline
\hline
\end{tabular}
\end{center}

%*******************************

\newpage  
As we see, about 2570 reconstructed $B_s^0$ mesons are expected for $10^4
pb^{-1}$ integrated luminosity from $B_s^0 \to D_s^- a_1^+$ channel when
both of $D_s^- \to \phi \pi^-$ and $D_s^- \to K^{*0} K^-$ modes are used 
for $D_s^-$ reconstruction. To them we should add 3640 $B_s^0$ mesons from
$B_s^0 \to D_s^- \pi^+$ channel \cite{8}. So in total we expect 6210
reconstructed $B_s^0$ mesons per $10^4 pb^{-1}$ integrated luminosity when 
all as yet considered decay modes are used.

\section{Peak amplifier}
To extract the oscillation frequency, from M.C. or experimental asymmetry
distributions, the so called amplitude fit method is useful \cite{11}. Here
we describe some refinements of this method.

In the amplitude fit method an asymmetry distribution $A(t)$ is fitted with
the cosine function $A_{fit}\cos {(x_st/\tau)}$ in which $x_s$ is fixed and
$A_{fit}$ is the only free parameter. Repeating the fit for different values
of $x_s$, we get $A_{fit}(x_s)$ distribution. This distribution is peaked
at $x_s$ which corresponds to the true value of the oscillation frequency.

The peak position in the $A_{fit}(x_s)$ distribution can be found with the 
help of the recently suggested "quantum" peak finding algorithm \cite{12}.
The idea of this algorithm is based on the property of small quantum balls
to penetrate narrow enough obstacles. So if such a ball is placed on the
edge of some potential wall it will find its way down to the potential wall
bottom even if the potential wall is distorted by statistical fluctuations.

Let us introduce instead of continuous $x_s$ some discrete parameter $i$,
say through $N_i=A_{fit}(i/2)$. The transformation $A_{fit}(x_s) \to
u(x_s)$, which we call peak amplifier, is defined for selected discrete
values of $x_s$ as follows
\begin{eqnarray}
u(i/2) \equiv u_i \; , \; u_{i+1}=\frac{P_{i,i+1}}{P_{i+1,i}}u_i \; , \;
\sum u_i =1 \; .
\label{eq2} \end{eqnarray} \noindent
And $P_{i,i\pm 1}$ transition probabilities are determined by the initial
$N_i$ spectrum \cite{12}
\begin{eqnarray}
P_{i,i \pm 1} = A_i \, \sum_{k=1}^{2} \exp { \left [ \frac{N_{i \pm k}
-N_i}{\sqrt{\sigma^2_{i \pm k}+\sigma^2_i}} \right ] } \hspace*{2mm} ,
\label{eq3} \end{eqnarray}
\noindent
$A_i$ normalization constant being defined from the $P_{i,i-1}+P_{i,i+1}=1$
condition. $\sigma_i$ is a standard deviation (error) of $N_i$ as determined
by the cosine fit.

If now we apply this peak amplifier to the data after the amplitude fit
we get the probability distributions shown on Fig. 5 (for $x_s=30$) and
on Fig. 6 (for $x_s=45$). As we see, the peak
amplifier enables a clear determination of $x_s$ from the amplitude fit
spectrum.

\section {$x_s$ sensitivity range}
To estimate the ATLAS sensitivity range for the $x_s$ measurement, 
the analogous procedure was used as in \cite{8}. 

Amplitude fit is applied to the asymmetry distribution generated by 
Monte-Carlo program. The input parameters of this program, such as signal 
to background ratio, $B_s^0$ lifetime, proper-time resolution and dilution 
factors are the same as in \cite{8}, with the exception of the number of
signal events, which was taken to be 6210.

The amplitude fit spectrum is further transformed using the peak amplifier
transformation as described above. In the resulting $u(x_s)$ spectrum
the mean value of $x_s$ and its standard deviation is calculated considering
$u(x_s)$ as a probability density. The "experiment" is considered as 
successful if the measured $x_s$ value (mean value of $x_s$ according to 
the $u(x_s)$ distribution) is within two standard deviations from the
true $x_s$-value defined in the Monte-Carlo program.

For each $x_s$ point 1000 such "experiments" were generated and the fraction
of the successful "experiments" was calculated. The highest value of $x_s$,
for which this fraction is above 95\%, is considered as a sensitivity limit
for the ATLAS experiment. This limit was found to be about $x_s^{max}=42.5$.
This is almost the same number as found in \cite{8}. In fact the peak
amplifier method doesn't give a significant increase for the sensitivity
limit, but it allows a more accurate $x_s$ determination, 
as is indicated by Fig. 6, because the probability peak is much more narrow.

Fig. 7 shows the distribution of the $x_s$ values, found by the peak
amplifier method, for 1000 "experiments", generated with the "true" 
$x_s=42.5$.
 
\section{"Where is the beginning of the end that comes at the end of the
beginning?"}
So we are at the end of our investigation. Our main conclusions are:
\begin{itemize}
\item $D_s^- \to K^{*0}K^-$ mode enables a two fold increase in the signal
statistics for the $B_s^0 \to D_s^-a_1$  decay channel.
\item the ATLAS experiment can reach a sensitivity limit for $x_s$ as
high as $x_s^{max}=42.5$ with the certainty.
\item $D_s^- \to K^{*0}K^-$ mode can be used also for the $B_s \to D_s^-
\pi^+$ channel. If the same increase in signal statistics is assumed,
the total number of reconstructed $B_s$ events can reach $10^4$ per
$10^4 pb^{-1}$ integrated luminosity. According to estimates from \cite{8}
this will mean a sensitivity limit for $x_s$ about 46.
\end{itemize}

We began our story with the Black Square. Here is one more Black Square
image which illustrates the continues progress in the $B_s$-mixing studies.
\newpage
% !!!!!!!!!!!!!!!!!!!!!!!!!!!!!!!!!!!!!!!!!!!!!!!!!!!!!!!!!!
% escher4.ps is too big. I'm sending escher4.jpg instead.
\vspace*{7cm}
\centerline{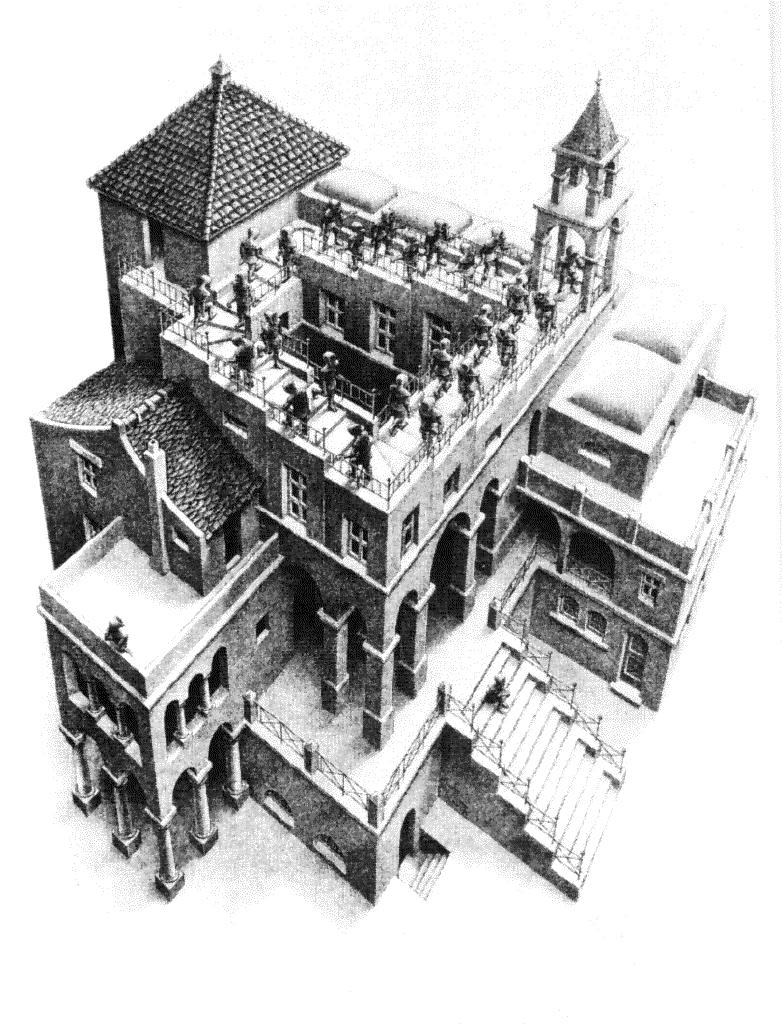 here}
\vspace*{7cm}
% !!!!!!!!!!!!!!!!!!!!!!!!!!!!!!!!!!!!!!!!!!!!!!!!!!!!!!!!!!
%
%\begin{figure}[htb]
%  \begin{center}
%    \mbox{\epsfig{figure=escher4.ps,%
%                        height=15.0cm}}
%  \end{center}
%\end{figure}

This picture leads to a question about a meaning of the scientific progress, 
or even to a more general question  \cite{13}
"What profit hath a man of all his labour which he taketh under the sun?".
We refrain to give any other comment about this things because \cite{14}
"Wo alle Worte zu wenig w\"{a}ren, dort ist jedes Wort zu viel". 

Nevertheless we don't want to end with only black and white images of the 
$B_s$-mixing related stuff. With the very great imagination you can catch
sight of the glorious full colour artist's view on the particle mixing
behind the Black Square (note the role of vertexing in emergence of this
dream):
\newpage
% !!!!!!!!!!!!!!!!!!!!!!!!!!!!!!!!!!!!!!!!!!!!!!!!!!!!!!!!!!
% tigers.ps is too big. I'm sending tigers.jpg instead.
\vspace*{7cm}
\centerline{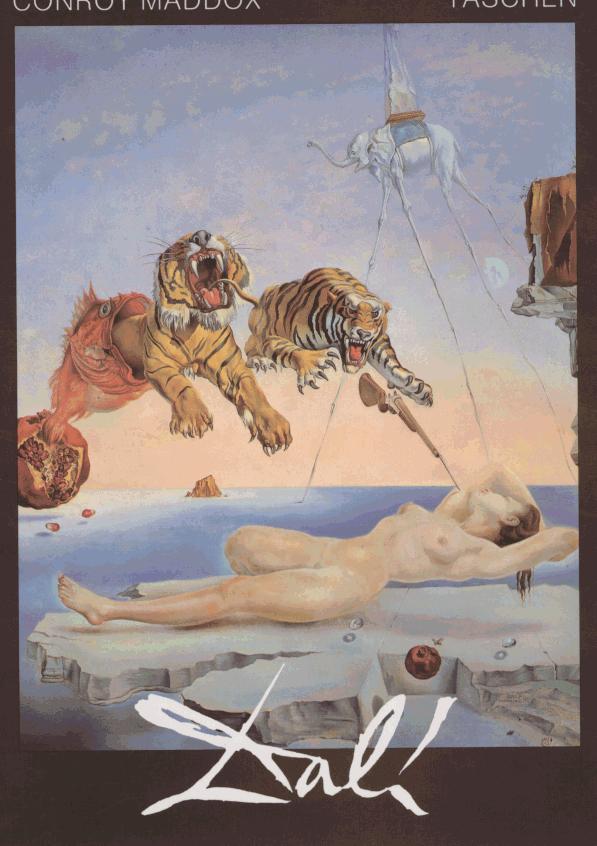 here}
\vspace*{7cm}
% !!!!!!!!!!!!!!!!!!!!!!!!!!!!!!!!!!!!!!!!!!!!!!!!!!!!!!!!!!
%
%\begin{figure}[htb]
%  \begin{center}
%    \mbox{\epsfig{figure=tigers.ps,%
%                        height=16.0cm}}
%  \end{center}
%\end{figure}
 
\newpage
\section*{Acknowledgments}
The authors are grateful to Paula Eerola and Szymon Gadomski 
for valuable comments. These comments were used in the text. 

\section*{Farewell} 
When we began this investigation our intention was to write a vivid
and joyful story about $B_s$-mixing, some mixture of science and art.
Only scientific framework appeared to us as too narrow to embrace the 
beauty of life, because \cite{15} "All things are full of labour; man
cannot utter it: the eye is not satisfied with seeing, nor the ear 
filled with hearing.".

Unfortunately Sasha Bannikov suddenly died at the end of last year and
we are forced to end this project without him.

Farewell Sasha! Let this article be a small thing that remains after you
in this world as your memory.

%********** Figures ************
\begin{figure}
\vspace*{2cm}
\epsfig
{file=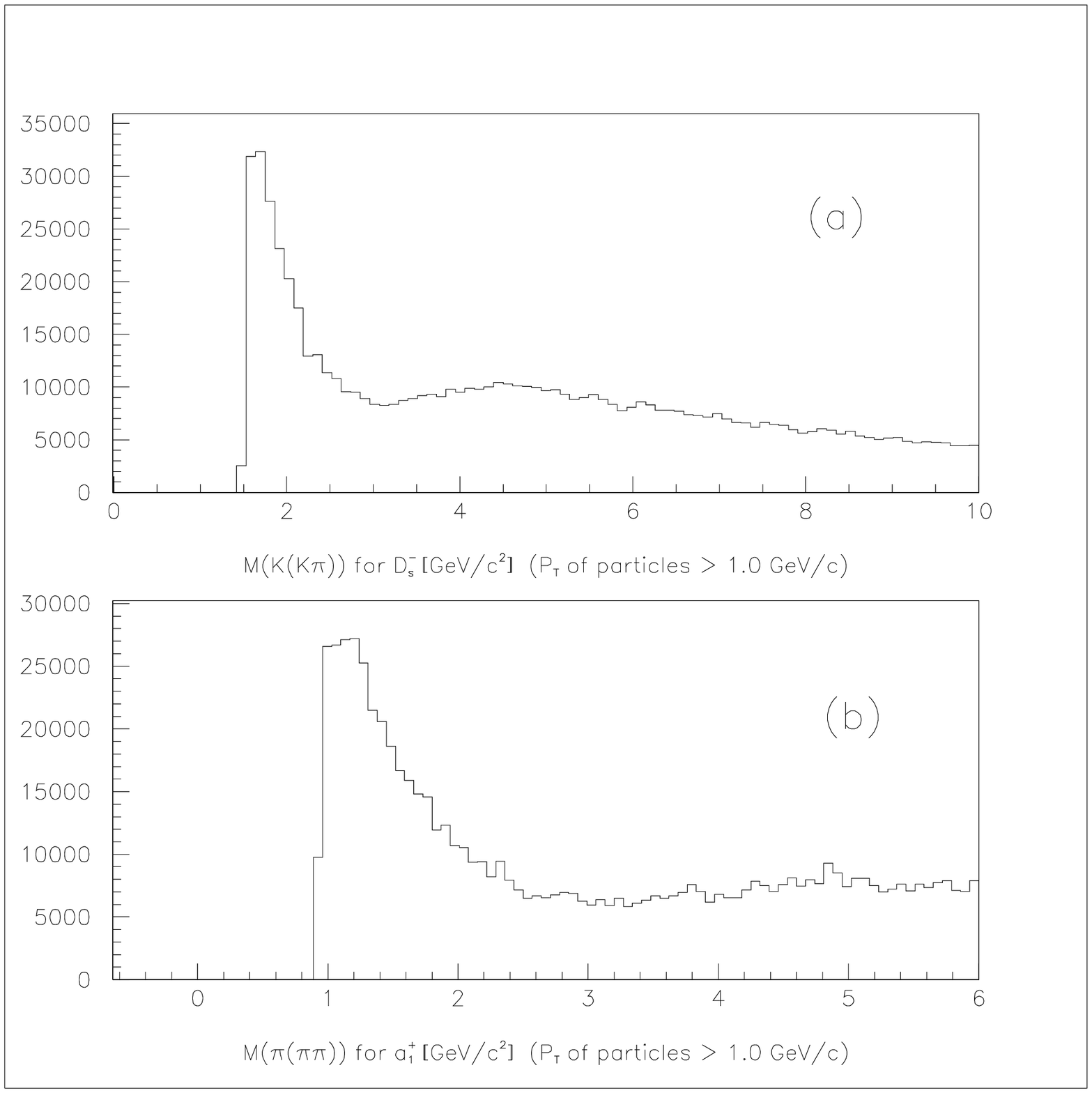,height=13cm,bbllx=40pt,bblly=70pt,bburx=625pt,bbury=550pt}
\caption{Invariant mass distributions of three charged particle combinations
in signal events,
assuming $2K+\pi$ (a) or $3\pi$ combination (b) .}
\end{figure}

\begin{figure}
\epsfig
{file=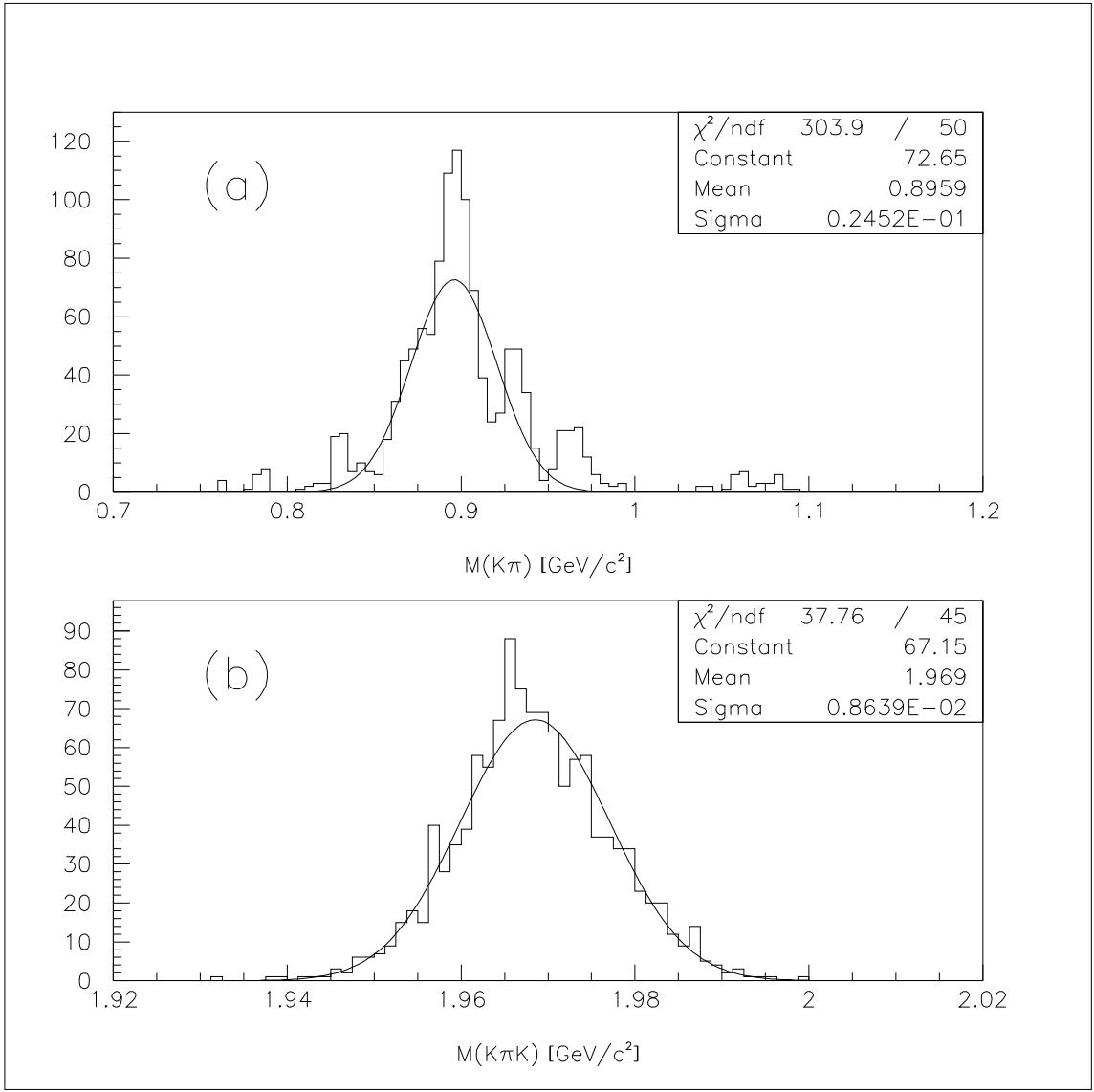,height=13cm,bbllx=40pt,bblly=70pt,bburx=625pt,bbury=550pt}
\caption{Invariant mass distributions
of reconstructed $K^{*0}$ and $D^-_s$ events.}
\end{figure}

\begin{figure}
\epsfig
{file=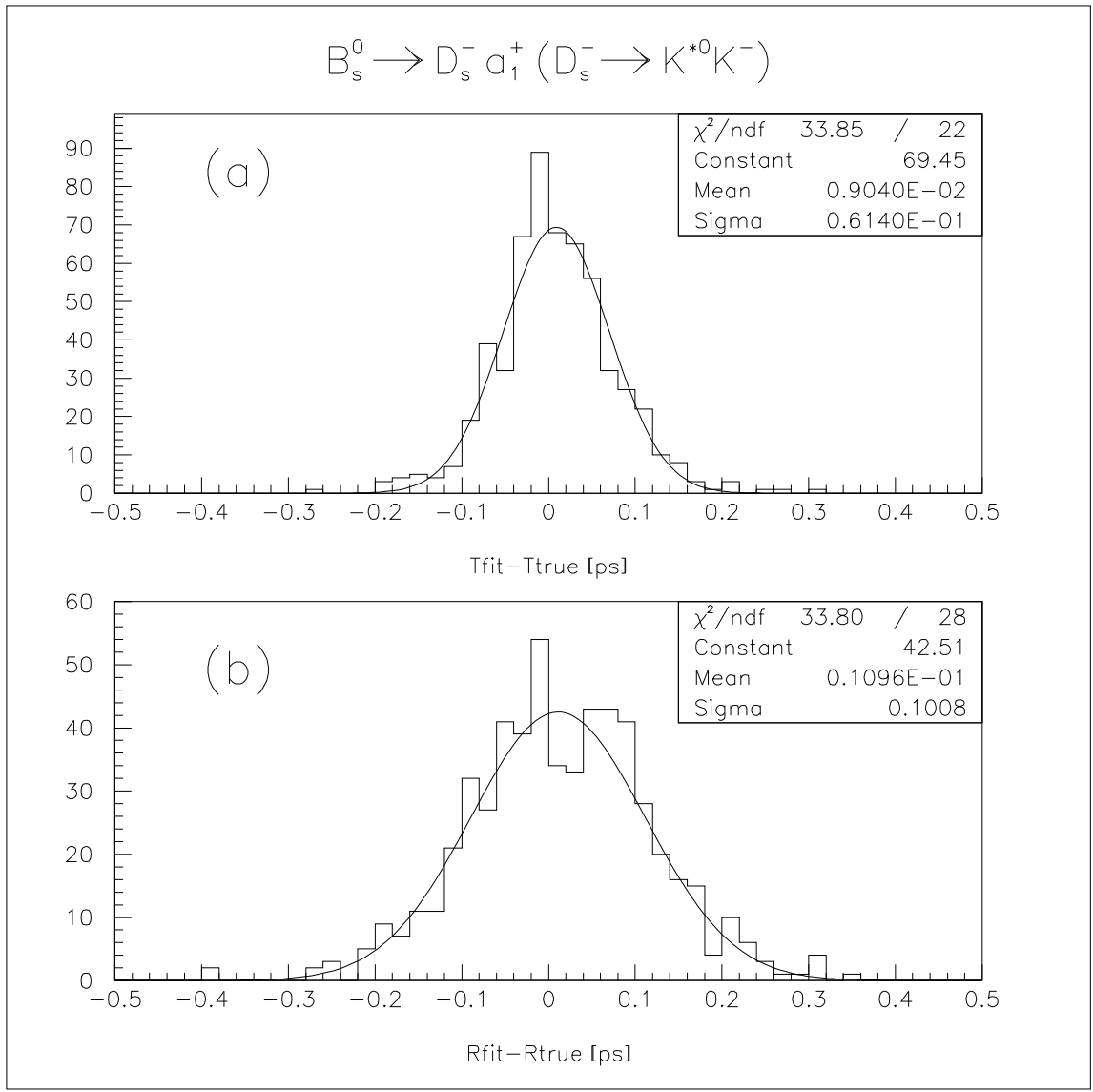,height=13cm,bbllx=40pt,bblly=70pt,bburx=625pt,bbury=550pt}
\caption{Proper time (a) and transverse radius (b) resolutions for the
reconstructed $B_s^0$ decay vertex.}
\end{figure}

\begin{figure}
\epsfig
{file=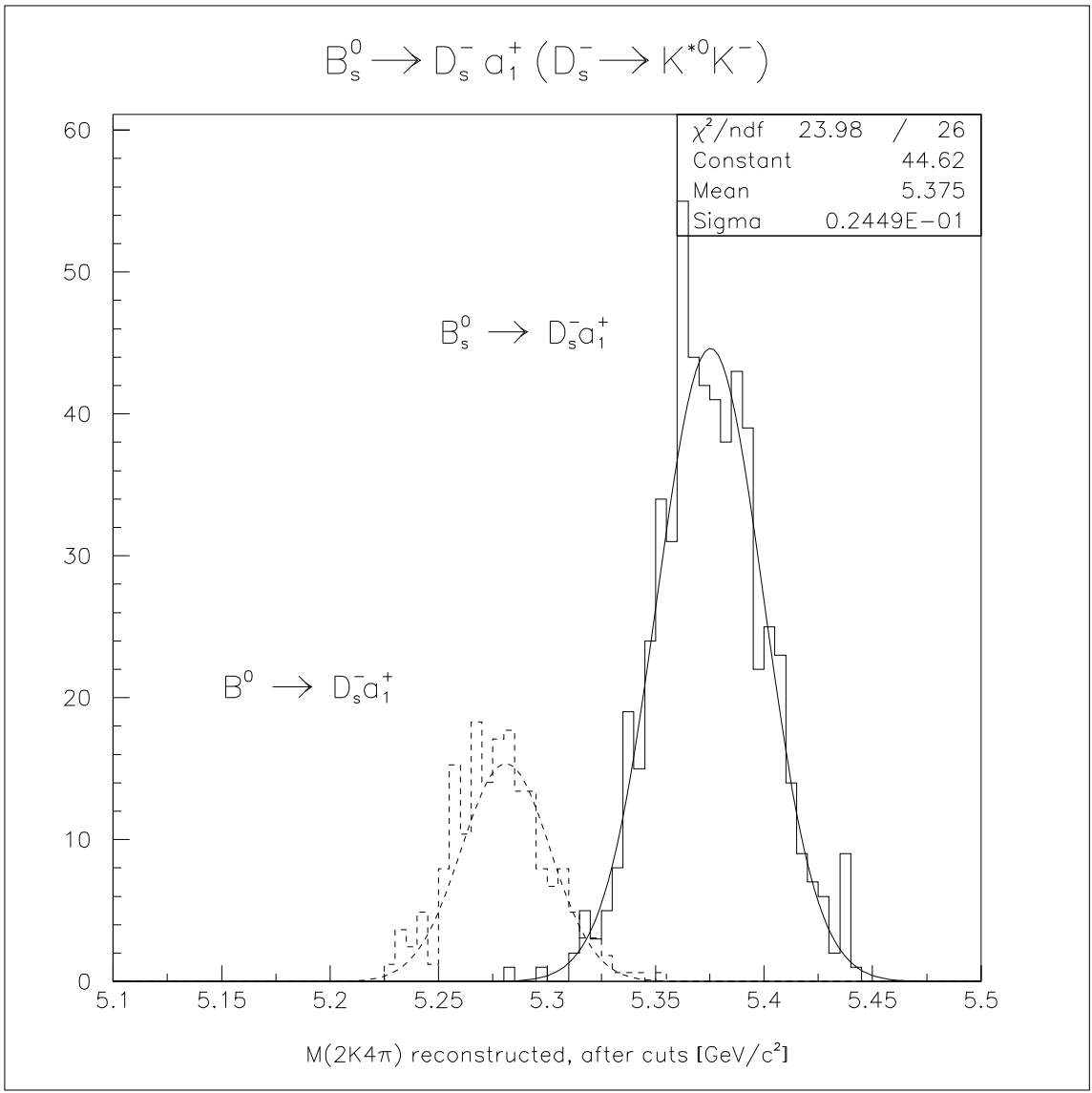,height=13cm,bbllx=40pt,bblly=70pt,bburx=625pt,bbury=550pt}
\caption{Six particle invariant mass distribution corresponding to the
$B_s^0$ meson. Dashed line - expected upper limit for background from
$B^0$ decay.}
\end{figure}

\begin{figure}
\epsfig
{file=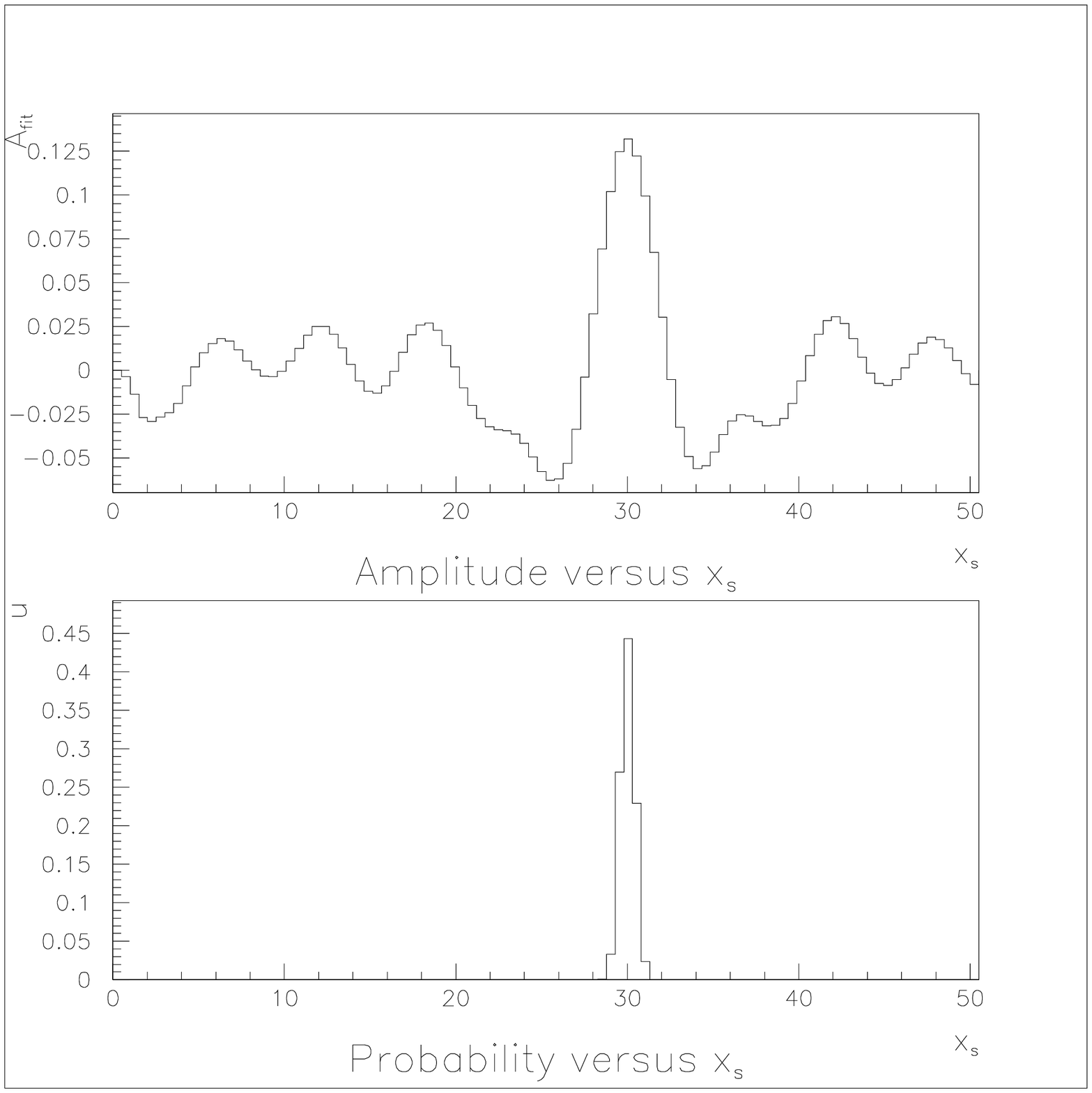,height=13cm,bbllx=40pt,bblly=70pt,bburx=625pt,bbury=550pt}
\caption{amplitude $A(x_s)$ and probability $u(x_s)$ for $x_s=30$.}
\end{figure}

\begin{figure}
\epsfig
{file=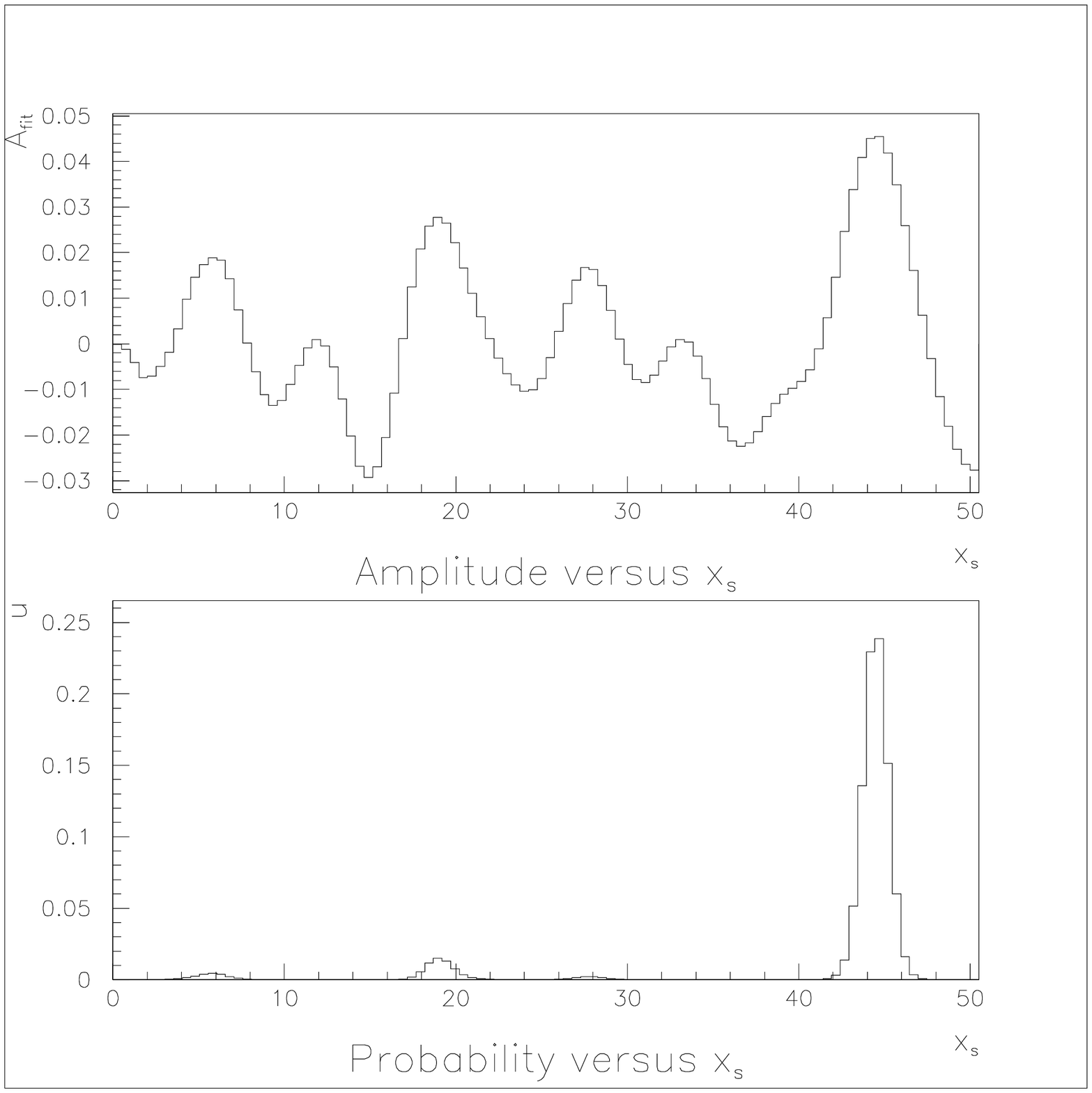,height=13cm,bbllx=40pt,bblly=70pt,bburx=625pt,bbury=550pt}
\caption{amplitude $A(x_s)$ and probability $u(x_s)$ for $x_s=45$.}
\end{figure}

\begin{figure}
\epsfig
{file=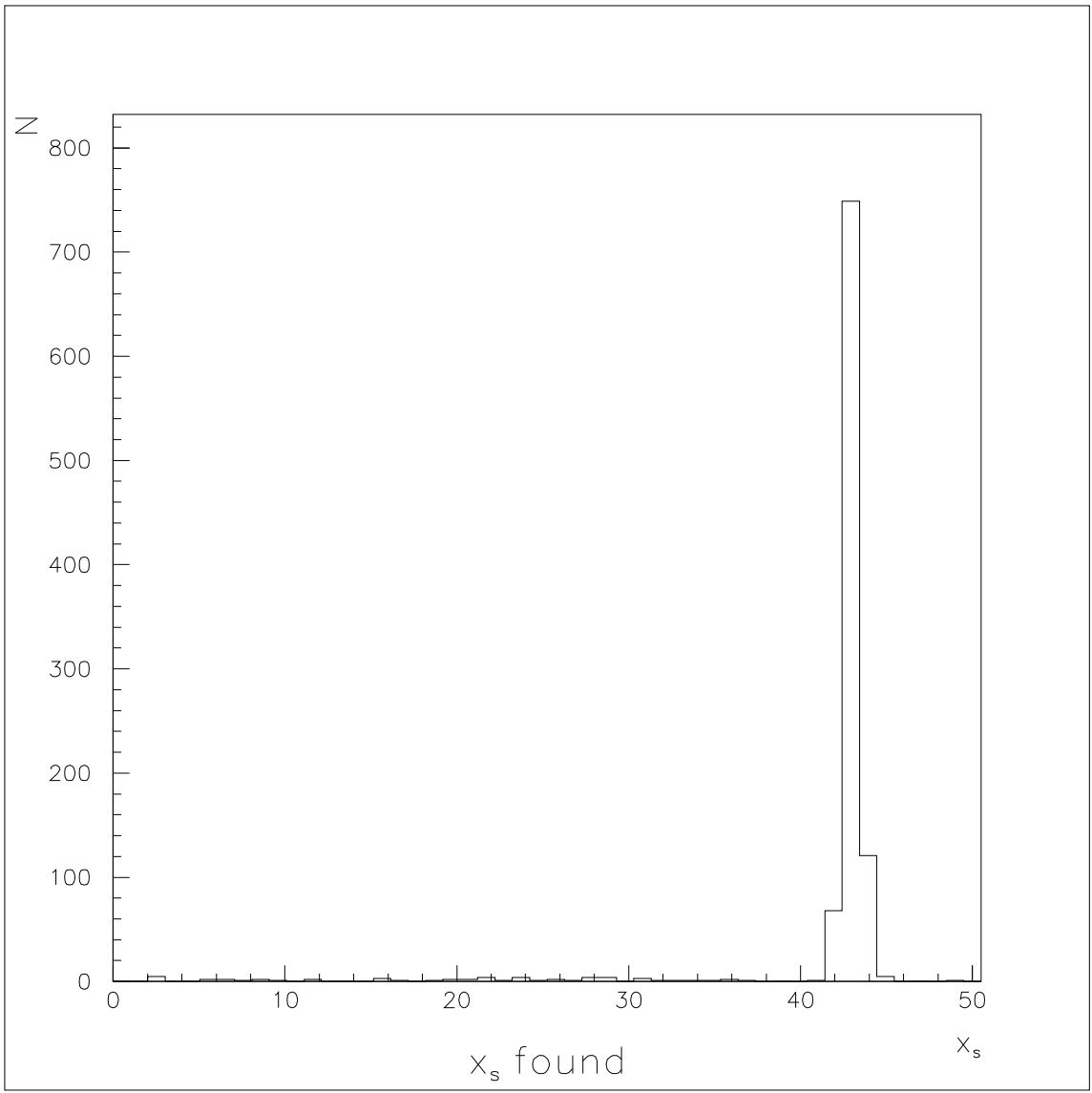,height=13cm,bbllx=40pt,bblly=70pt,bburx=625pt,bbury=550pt}
\caption{The distribution of the $x_s$ values, found by the peak
amplifier method, for 1000 "experiments", generated with the $x_s=42.5$.}
\end{figure}

\end{document}